\documentclass[12pt, final]{l4dc2024}
\usepackage{xz_style}
\usepackage{listings}
\lstdefinestyle{mystyle}{
    language=Python,
    basicstyle=\ttfamily\small,
    linewidth=\textwidth,
    keywordstyle=\color{blue},
    commentstyle=\color{green},
    stringstyle=\color{red},
    frame=single, 
    numbers=none, 
    numberstyle=\tiny\color{gray}, 
    stepnumber=1, 
    numbersep=10pt, 
    backgroundcolor=\color{white}, 
    showspaces=false, 
    showstringspaces=false, 
    showtabs=false, 
    tabsize=2, 
    captionpos=b, 
    breaklines=true, 
    breakatwhitespace=true, 
    commentstyle=\color{gray}\ttfamily, 
    morekeywords={*,...}, 
}
\usepackage{array, threeparttable} 

\title[Controlgym: Large-Scale Control Environments]{Controlgym: Large-Scale Control Environments for Benchmarking Reinforcement Learning Algorithms}
\usepackage{times}

\author{%
\Name{Xiangyuan Zhang} \Email{xz7@illinois.edu}\\
\Name{Weichao Mao} \Email{weichao2@illinois.edu}\\
\addr ECE \& CSL, University of Illinois Urbana-Champaign \\
\Name{Saviz Mowlavi} \Email{mowlavi@merl.com}\\
\Name{Mouhacine Benosman} \Email{benosman@merl.com}\\
\addr Mitsubishi Electric Research Laboratories \\
\Name{Tamer Ba\c{s}ar} \Email{basar1@illinois.edu}\\
\addr ECE \& CSL, University of Illinois Urbana-Champaign \\
}

\begin{document}
\maketitle

\begin{figure}[b]
	\centering
	\includegraphics[height = 4.3cm, width = 0.95\linewidth]{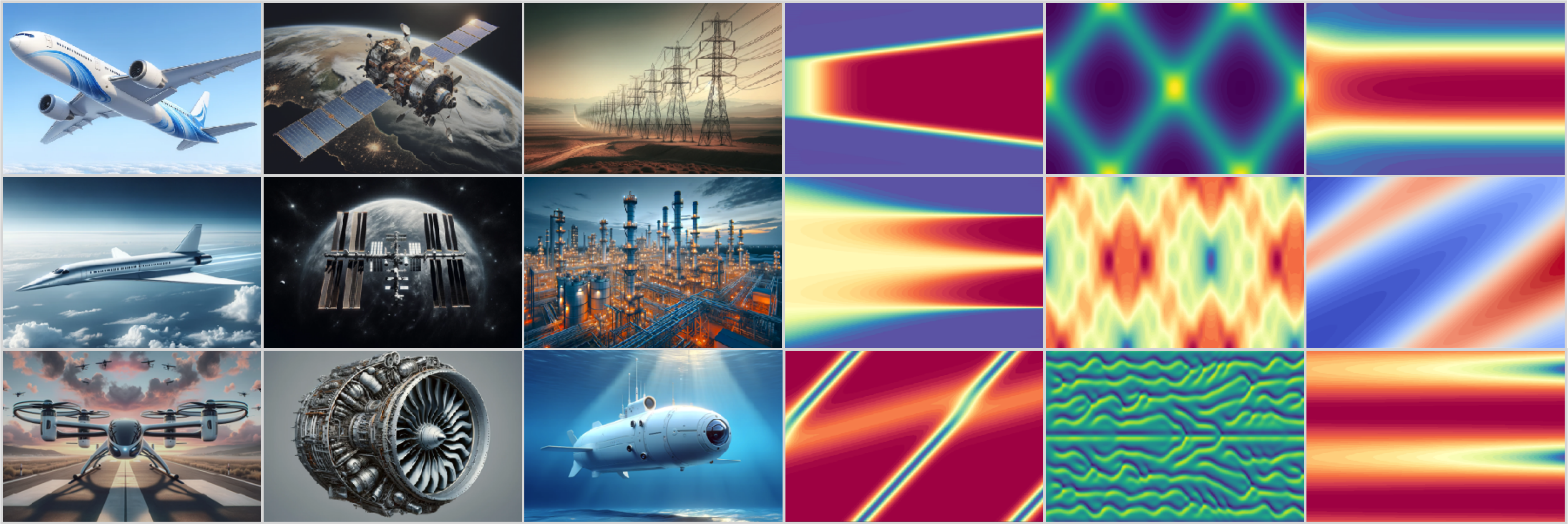}
	\vspace{-0.5em}
	\caption{Environments included in this work, motivated by industrial control applications }
	\label{fig:gallery}
\end{figure}

\begin{abstract}%
We introduce \texttt{controlgym}, a library of thirty-six industrial control settings, and ten infinite-dimensional partial differential equation (PDE)-based control problems. Integrated within the OpenAI Gym/Gymnasium (Gym) framework, \texttt{controlgym} allows direct applications of standard reinforcement learning (RL) algorithms like stable-baselines3. Our control environments complement those in Gym with continuous, unbounded action and observation spaces, motivated by real-world control applications. Moreover, the PDE control environments uniquely allow the users to extend the state dimensionality of the system to infinity while preserving the intrinsic dynamics. This feature is crucial for evaluating the scalability of RL algorithms for control. This project serves the learning for dynamics \& control (L4DC) community, aiming to explore key questions: the convergence of RL algorithms in learning control policies; the stability and robustness issues of learning-based controllers; and the scalability of RL algorithms to high- and potentially infinite-dimensional systems. We open-source the \texttt{controlgym} project at \url{https://github.com/xiangyuan-zhang/controlgym}. %
\end{abstract}


\section{Introduction}\label{sec:intro}
The intersection of machine learning (ML), reinforcement learning (RL), and control theory has garnered significant attention in recent years, giving rise to the learning for dynamics \& control (L4DC) research community \citep{recht2019tour, vamvoudakis2021handbook, brunke2021safe, hu2023toward}. L4DC has the naturally driven mission to unlock the power of learning-based methods for control and establish a rigorous theoretical foundation \citep{L4DC}. This mission could only be fulfilled with joint forces and close collaboration between theorists and practitioners from ML, control theory, and optimization. 

Theorists are keen to validate their algorithms and theories in real-world scenarios but encounter challenges with OpenAI Gym/Gymnasium (Gym) environments \citep{brockman2016openai, towers_gymnasium_2023}. Specifically, most Gym environments feature highly nonlinear dynamics, often involving contacts, and offer very limited parameter customization options, making them ill-suited testbeds for control theory research. Meanwhile, control textbook examples lack the complexity for cutting-edge ML/RL research that prioritizes efficiency and scalability.

To address these requirements, we introduce \texttt{controlgym}, a lightweight and versatile Python library that offers a spectrum of environments spanning from linear systems to chaotic, large-scale systems governed by partial differential equations (PDEs). Specifically, \texttt{controlgym} features thirty-six linear industrial control environments, encompassing sectors like aerospace, cyber-physical systems, ground and underwater vehicles, and power systems. Additionally, \texttt{controlgym} includes ten large-scale control environments governed by fundamental PDEs in fluid dynamics and physics. These PDEs are discretized in space by custom solvers, yielding user-tunable state-space dimensions without affecting the dynamics of the environment, a key aspect for assessing the scalability of RL algorithms. All environments comply with Gym and support standard RL algorithms \citep{sutton2000policy, kakade2002natural, schulman2015trust, schulman2017proximal, mnih2016asynchronous, sutton2018reinforcement}, e.g., as seen in stable-baselines3 \citep{raffin2021stable}.

Our primary contribution is the introduction of a diverse array of control environments characterized by continuous and unbounded action-observation spaces, designed for large-scale systems. These environments, detailed in Tables \ref{table:control} and \ref{table:pde}, enhance Gym's collection and are highly customizable to support theoretical advancement in L4DC.  For example, users can manipulate the open-loop dynamics of PDEs by adjusting physical parameters, with explicit formulas relating parameters and eigenvalues available in linear PDE environments (cf., Section \ref{sec:pde_eigens}). Moreover, our PDE environments uniquely allow the users to extend system dimensionality to infinity while preserving the intrinsic dynamics. The PDE solvers implemented to power \texttt{controlgym} are innovative, employing state-of-the-art schemes with exponential spatial convergence and high-order temporal accuracy masked behind a user-friendly discrete-time state-space formulation. Specifically for linear PDE environments, we have developed novel state-space models to evolve the PDE dynamics.

Leveraging its strengths, \texttt{controlgym} is a testbed for exploring three essential aspects of applying RL to continuous control. First, it aims to probe whether RL algorithms can consistently converge in learning control policies. Second, it examines the stability and robustness of the policy and training process, motivated by real-world safety-critical applications. Lastly, it assesses the scalability of RL algorithms in high-dimensional and potentially infinite-dimensional systems. With \texttt{controlgym}, we bridge the theoretical development and practical applicability of L4DC by providing a research platform that supports the establishment of a rigorous foundation. Initial deployments of \texttt{controlgym} include RL for PDE control \citep{zhang2024policy, botteghi2024parametric} and toward a foundational control transformer \citep{zhang2024decision}.

\vspace{0.5em}
\noindent\textbf{Related works. } The COMP$l_{e}$ib project \citep{leibfritz2004compleib1, leibfritz2004compleib2} pioneered in offering standard control tasks, as MATLAB files, for analyzing model-based control algorithms. In the era of ML and RL, Gym \citep{brockman2016openai, towers_gymnasium_2023} has become the standard platform for developing and benchmarking RL algorithms for continuous control, offering a variety of environments such as cart pole, inverted pendulum, and robotic tasks powered by Mujoco \citep{todorov2012mujoco}. Numerous follow-up projects that implement RL algorithms on Gym environments include rllab/garage \citep{duan2016benchmarking}, RLlib \citep{liang2018rllib}, dm$\_$control  \citep{tunyasuvunakool2020}, deluca \citep{gradu2020deluca}, stable-baselines3 \citep{raffin2021stable}, safe-control-gym \citep{brunke2021safe}, realworldrl-suite \citep{dulac2020empirical}, tianshou \citep{tianshou}, and TorchRL \citep{Bou_TorchRL_A_data-driven_2023}. Very recently, HydroGym \citep{hydrogym} provided fluid dynamics environments for testing RL algorithms for flow control. 

Complying with Gym's framework, we offer a spectrum of control environments designed to support the foundational theoretical developments in RL for linear optimal control \citep{fazel2018global, bu2019lqr, tu2019gap, mohammadi2019convergence, yang2019provably, dean2020sample, malik2020derivative, furieri2019learning, simchowitz2020naive, simchowitz2020improper, hambly2020policy, chen2021black, perdomo2021stabilizing, li2019distributed, zhao2021primal, jansch2022policy, ozaslan2022computing, ju2022model, lale2022reinforcement, duan2022optimization, zhang2023revisiting, zheng2021analysis, tsiamis2022statistical, ziemann2022policy, duan2023optimization}, linear robust control and dynamic games \citep{agarwal2019online, gravell2019learning, bu2019global, zhang2019policy, zhang2019policymixed, yang2020h, zhang2021provably, zhang2020rarl, zhang2021derivative, keivan2022model, guo2022global,  cui2023reinforcement}, estimation and filtering \citep{umenberger2022globally, zhang2023learning, zhang2023global, zhang2023ifac}, and PDE control \citep{pan2018reinforcement, bucci2019control, liu2021physics, degrave2022magnetic, zeng2022data, vignon2023recent, mowlavi2023reinforcement, werner2023learning, peitz2023distributed}. 

\vspace{0.5em}
\noindent\textbf{Notations. }In the paper, we follow the list of notations in the following table.
\vspace{-0.2em}
\begin{table}[h!]
\setlength\defaultaddspace{0.0ex}
\centering
\small
\begin{threeparttable}
    \begin{tabular}{@{}cm{13cm}@{}}
    \toprule
    \textbf{Notation} & \textbf{Description} \tabularnewline
    \midrule
$s_k$, $a_k$ & system state and control input/action at discrete time $k$, respectively \tabularnewline
$w_k$, $v_k$ & stochastic noise or deterministic uncertainty at discrete time $k$ \tabularnewline
$n_s$, $n_a$, $n_y$ & dimensionalities of state, action, and observation, respectively \tabularnewline
$\Omega$, $L$, $x$, $u$ & physical domain, domain length, spatial coordinates, and spatial field of PDE, respectively \tabularnewline
$\Phi_i(x)$ & time-invariant forcing support function for the $i^{th}$ control input $a_i$ \tabularnewline 
$\Delta t, dt$ & sampling time and numerical integration time step, respectively \tabularnewline 
$K$ & total number of discrete time steps \tabularnewline 
$W$ & discrete Fourier transform (DFT) matrix \tabularnewline 
$I$, $0$ & Identity and zero matrices of appropriate dimensions, respectively \tabularnewline
$c$ & velocity in the convection-diffusion-reaction (CDR) and wave equations \tabularnewline 
$\nu$ & diffusivity constant in CDR, Burgers', Fisher, Allen-Cahn, and Cahn-Hilliard equations \tabularnewline 
$r$ & reaction constant in CDR and Fisher equations \tabularnewline
$\hbar, m$ & Planck constant and particle mass, respectively, in the Schrödinger equation \tabularnewline
$V$ & potential constant in Schrödinger and Allen-Cahn equations \tabularnewline
$\Gamma$ & surface tension constant in the Cahn-Hilliard equation \tabularnewline
$\psi$ & local change rate of $u(x, t)$ in the wave equation \tabularnewline
$\xi$, $\eta$ & real and imaginary parts of $u(x, t)$ in the Schrödinger equation, respectively \tabularnewline
    \bottomrule
    \end{tabular}%
\end{threeparttable}
\nonumber
\vspace{-0.5em}
\end{table}%

\section{Control Environments}
\subsection{Linear Control Environments}
\begin{table}[t]
\centering
\footnotesize
\caption{List of the linear control environments in \texttt{controlgym}}
\vspace{0.3em}
\label{table:control}
\begin{threeparttable}
    \begin{tabular}{ccccm{3.0cm}|ccccm{4.3cm}}
    \toprule
    \textbf{ID} & $n_s$ & $n_a$ & $n_y$ & \textbf{Task}  & \textbf{ID} & $n_s$ & $n_a$ & $n_y$ & \textbf{Task} \tabularnewline
    \midrule
ac1 & 5 & 3 & 3 & aircraft & cm5 & 480 & 1 & 2 & cable-mass model \tabularnewline  
ac2 & 5 & 3 & 3 & aircraft & dis1 & 8 & 4 & 4 & decentralized system \tabularnewline  
ac3 &4 & 1 & 2 & aircraft & dis2 & 4 & 2 & 2 & decentralized system\tabularnewline  
ac4  & 9 & 1 & 2 & aircraft & dlr & 40 & 2 & 2 & space structure\tabularnewline  
ac5  & 9 & 1 & 5 & aircraft & he1 & 4 & 2 & 1 & helicopter \tabularnewline  
ac6  & 10 & 4 & 5 & aircraft & he2 & 8 & 4 & 6 & helicopter \tabularnewline  
ac7  & 55 & 2 & 2 & aircraft & he3 & 8 & 4 & 6 & helicopter \tabularnewline  
ac8  & 4 & 3 & 4 & aircraft & he4 & 8 & 4 & 2 & helicopter \tabularnewline  
ac9  & 40 & 3 & 4 & aircraft & he5 & 20 & 4 & 6 & helicopter  \tabularnewline  
ac10  & 10 & 2 & 2 & aircraft & he6 & 20 & 4 & 6 & helicopter \tabularnewline  
bdt1  & 11 & 3 & 3 & distillation tower & iss & 270 & 3 & 3 & International Space Station \tabularnewline  
bdt2  & 82 & 4 & 4 & distillation tower & je1 & 30 & 3 & 5 & jet engine  \tabularnewline  
 cbm & 348 & 1 & 1 & clamped beam model & je2 & 24 & 3 & 6 & jet engine \tabularnewline  
cdp & 120 & 2 & 2 & CD player & lah & 48 & 1 & 1 & L.A. University Hospital \tabularnewline  
cm1  & 20 & 1 & 2 & cable-mass model & pas & 5 & 1 & 3 & piezoelectric bimorph actuator \tabularnewline  
cm2  & 60 & 1 & 2 & cable-mass model & psm & 7 & 2 & 3 & power system  \tabularnewline  
cm3  & 120 & 1 & 2 & cable-mass model & rea & 8 & 1 & 1 & chemical reactor \tabularnewline  
cm4 &240 & 1 & 2 & cable-mass model  & umv & 8 & 2 & 2 & underwater vehicle
 \tabularnewline
   \bottomrule
\end{tabular}\par 
\end{threeparttable}
\end{table}%

We incorporate $36$ linear control environments from various industries, as detailed in Table \ref{table:control}. We select and organize these continuous-time linear systems from the pioneering COMP$l_{e}$ib project \citep{leibfritz2004compleib1, leibfritz2004compleib2}, and provide them as standard Gym environments. These environments span control applications ranging from aircraft, helicopters, jet engines, reactor models, decentralized cyber-physical systems, binary distillation towers, ground and underwater autonomous vehicles, power systems, compact disk (CD) players, and large space structures. Additionally, the scope of our environments extends to control problems within projects such as the International Space Station and the Los Angeles Hospital.

With the user-selected sampling time $\Delta t$, we assume the control input is constant over each $\Delta t$ and generate the discrete-time system dynamics as 
\begin{align*}
	s_{k+1} &= As_k + B_1w_k + B_2a_k, \\
	z_{k} &= C_1s_k + D_{11}w_k + D_{12}a_k, \\
	y_{k} &= Cs_k + D_{21}w_k,
\end{align*}
where $s_k \in \RR^{n_s}$ is the state, $a_k$ is the control/action input, $w_k$ is the disturbance input that could be either stochastic or adversarial, $z_k \in \RR^{n_z}$ is the output, $y_k \in \RR^{n_y}$ is the observation, and $A$, $B_1$, $B_2$, $C_1$, $C$, $D_{11}$, $D_{12}$, $D_{21}$ are the discretized system matrices with appropriate dimensions. These linear control environments are directly applicable to support theoretical research of RL for the fundamental linear control, games, and estimation tasks (cf., Section \ref{sec:intro}).

\vspace{0.5em}
\noindent\textbf{Linear control objectives. }For each linear control task in Table \ref{table:control}, we define a regulation task whose primary objective is to steer the system's dynamics toward the zero vector. The reward function (to be maximized) is formulated as the negative sum of the linear-quadratic (LQ) stage cost 
\begin{align*}
	\cJ(a_k) = - \EE\Big\{\sum_{k=0}^{K-1} (s_k^{\top}Qs_k + a_k^{\top}Ra_k + 2s_k^{\top}Sa_k)\Big\},
\end{align*} 
where $Q = C_1^{\top}C_1$, $R = D_{12}^{\top}D_{12}$, and $S = C_1^{\top}D_{12}$ aim to balance regulation performance and control efforts, and $K$ is the total number of discrete time steps.

\subsection{PDE Control Environments}
\begin{figure}[h]
	\centering
	\includegraphics[width = 0.9\linewidth]{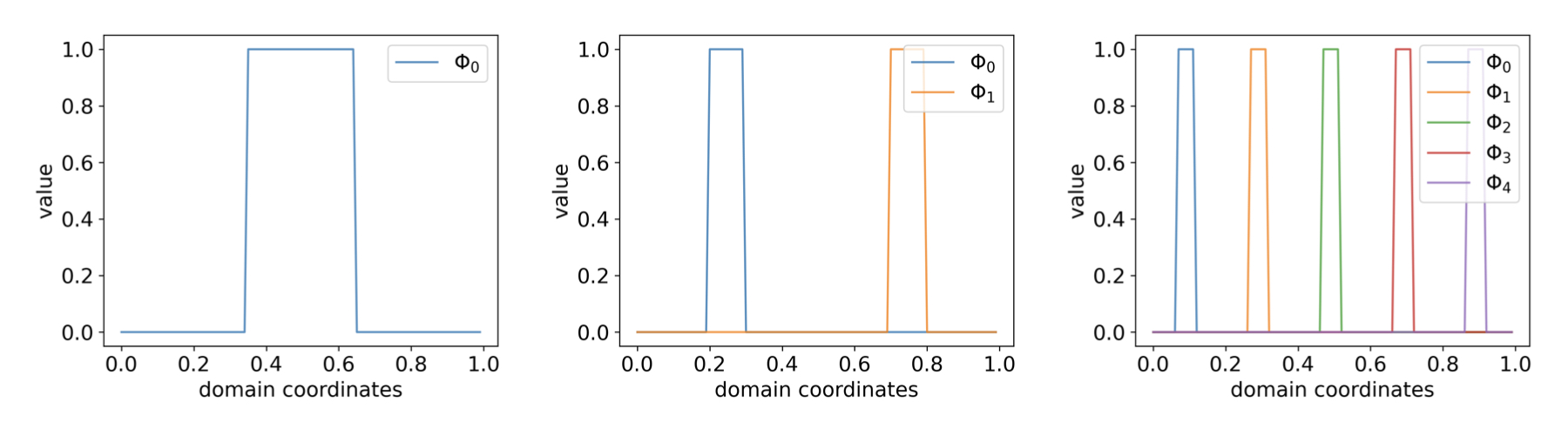}
	\caption{Illustration of how distributed control inputs influence the dynamics of a PDE through forcing support functions $\Phi_j$, taking $\Omega = [0,1]$ as an example. Left: The forcing support function corresponding to a single control input is depicted, with its width, a tunable parameter, set to $0.3$. This represents a control input that uniformly affects state components spanning the middle $30\%$ of the physical domain. Middle: The forcing support functions corresponding to two control inputs are shown. They are spaced equidistantly from one another, and each has a width of $0.1$ so that each control input uniformly affects state components spanning $10\%$ of the physical domain. Right: The forcing support functions corresponding to five control inputs are shown, each with a width of $0.05$ uniformly affecting state components spanning $5\%$ of the physical domain.}
	\label{fig:control_sup}
\end{figure}

\begin{table}[t]
\centering
\caption{List of PDE control environments in \texttt{controlgym}}
\label{table:pde}
\vspace{0.2em}
\begin{threeparttable}
\begin{tabular}{ccccm{2cm}}
    \toprule
    \textbf{ID} & $n_s$ & $n_a$ & $n_y$ & \textbf{Linearity}\tabularnewline
    \midrule
    	convection\_diffusion\_reaction & $(50, +\infty)$ & $(1, +\infty)$ & $(1, +\infty)$ & Linear  \tabularnewline  
wave & $(100, +\infty)$ & $(1, +\infty)$ & $(1, +\infty)$ & Linear \tabularnewline   
schrodinger & $(100, +\infty)$ & $(1, +\infty)$ & $(1, +\infty)$ & Linear \tabularnewline
burgers & $(50, +\infty)$ & $(1, +\infty)$ & $(1, +\infty)$ & Nonlinear \tabularnewline 
kuramoto\_sivashinsky & $(200, +\infty)$ & $(1, +\infty)$ & $(1, +\infty)$  & Nonlinear \tabularnewline  
fisher & $(50, +\infty)$ & $(1, +\infty)$ & $(1, +\infty)$ & Nonlinear \tabularnewline 
allen\_cahn & $(50, +\infty)$ & $(1, +\infty)$ & $(1, +\infty)$ & Nonlinear \tabularnewline
korteweg\_de\_vries & $(200, +\infty)$ & $(1, +\infty)$ & $(1, +\infty)$ & Nonlinear \tabularnewline    
cahn\_hilliard & $(50, +\infty)$ & $(1, +\infty)$ & $(1, +\infty)$ & Nonlinear \tabularnewline 
ginzburg\_landau & $(50, +\infty)$ & $(1, +\infty)$ & $(1, +\infty)$  & Nonlinear\tabularnewline  
    \bottomrule
    \end{tabular}\par 
\end{threeparttable}
\end{table}%
In this section, we describe one-dimensional PDE control environments with periodic boundary conditions and spatially distributed control inputs. We first define a spatial domain $\Omega = [0, L] \subset \mathbb{R}$ and a continuous field $u(x, t):\Omega \times \mathbb{R}^+ \rightarrow \mathbb{R}$, where $x$ and $t$ represent spatial and temporal coordinates, respectively, and $L$ is the length of the domain. Each PDE control task listed in Table \ref{table:pde} then takes the general continuous form
\begin{align}\label{eqn:general_pde}
	\frac{\partial u}{\partial t} - \mathcal{F} \left( \frac{\partial u}{\partial x}, \frac{\partial^2 u}{\partial x^2}, \dots \right) =  a(x, t),
\end{align}
where $\mathcal{F}$ is a linear or nonlinear differential operator (see Sections \ref{sec:cdr}-\ref{sec:ginzburg_landau} for specific definitions for each PDE) that contains spatial derivatives of various orders and depends on various physical constants, and $a$ is a distributed control force defined as 
\begin{align}\label{eqn:distributed_control}
	a(x, t) = \sum_{j = 0}^{n_a-1} \Phi_j(x) a_j(t).
\end{align}
The control force consists of $n_a$ scalar control inputs $a_j(t)$, each acting over a specific subset of $\Omega$ defined by its corresponding forcing support function $\Phi_j(x)$, as illustrated in Figure \ref{fig:control_sup}. Such a control force can be used to model the addition of energy to the system or other external influences that affect the PDE dynamics. We use periodic boundary conditions in all of our PDE control tasks, meaning that $u$ and all its spatial derivatives are equal at both ends of the domain $\Omega$.

\vspace{0.5em}
\noindent\textbf{Discretization of space and time. } To solve the PDEs listed in Table \ref{table:pde}, we first need to discretize space and time in the continuous form \eqref{eqn:general_pde}. For a state dimension $n_s$ that is even and a sampling time $\Delta t \in \RR^+$, both selected by the user, we define a state vector $s_k \in \RR^{n_s}$  that contains the values of $u$ at $n_s$ equally-spaced points in $\Omega$ and at discrete time $k \in \NN$ corresponding to the simulation time $t = k \, \Delta t$. The total simulation time is $K \, \Delta t$, where $K \in \NN$ is an input parameter specifying the total number of discrete-time steps. We also assume that the scalar control inputs $a_i(t)$ are piecewise constant over each discrete-time step of duration $\Delta t$ so that they can be concatenated into a discrete-time vector $a_k \in \RR^{n_a}$ for all $k \in \{0, \cdots, K-1\}$.

\vspace{0.5em}
\noindent\textbf{Numerical solver for nonlinear PDEs. } After discretizing space and time, the dynamics of the nonlinear PDEs can be approximated by a discrete-time finite-dimensional nonlinear system
\begin{align}\label{eqn:nonlinear_discrete}
	s_{k+1} = f(s_k, a_k; w_k),
\end{align}
where $f:\RR^{n_s}\times\RR^{n_a}\to\RR^{n_s}$ is a time-invariant mapping contingent on the physical parameters of each specific PDE and forcing support functions $\Phi_i$, and $w_k$ is an optional stochastic process noise. To compute the mapping $f$, we numerically approximate the space and time derivatives in \eqref{eqn:general_pde} using, respectively, a pseudo-spectral method \citep{trefethen1996finite} and a fourth-order exponential time differencing Runge-Kutta (ETDRK4) scheme \citep{cox2002exponential,kassam2005fourth}. We then integrate numerically the dynamics over one discrete-time step of duration $\Delta t$ using an internal integration time step $dt$. The mapping $f$ is therefore not obtained explicitly; rather, its action is evaluated through a numerical integration loop. The sampling time $\Delta t$ may be selected as large as desired, but it should be an integer multiple of the integration time step $dt$.

In general, one should choose $n_s$ to be sufficiently large and $dt$ to be sufficiently small to ensure the accuracy of the discretized system \eqref{eqn:nonlinear_discrete}. Due to the exponential convergence rate of the pseudo-spectral method as well as the high convergence rate of the ETDRK4 scheme, $n_s$ larger than about 50 is sufficient in most cases, except the Korteweg de Vries and Kuramoto-Sivashinsky PDEs that require $n_s$ larger than about 200 for accurate solutions. For all PDEs, the presence of small-scale (i.e., high wavenumber) spatial features in the initial condition may necessitate higher values of $n_s$.

\vspace{0.5em}
\noindent\textbf{Explicit state-space model for linear PDEs.} After space and time discretization, the linear PDEs listed in Table \ref{table:pde} can be approximated by a discrete-time linear state-space model of the form
\begin{align}\label{eqn:linear_discrete}
	s_{k+1} = A s_k + B_2 a_k + w_k,
\end{align}
where $A \in \RR^{n_s \times n_s}$ is a time-invariant transition matrix contingent on the physical parameters of each specific PDE, $B_2 \in \RR^{n_s \times n_a}$ is a time-invariant control matrix, and $w_k \sim \cN(0, \Sigma_w)$ is an optional process noise. The $A$ matrix in \eqref{eqn:linear_discrete} is constructed from a spectral approximation of the space derivatives in \eqref{eqn:general_pde} combined with an analytical temporal integration of the continuous-time linear dynamics over one discrete-time step of duration $\Delta t$ (see Sections \ref{sec:cdr}-\ref{sec:schrodinger} for detailed treatments of each case). Contrary to the case of nonlinear PDEs where evaluating the mapping $f$ in \eqref{eqn:nonlinear_discrete} requires an internal numerical integration loop, the availability of matrix $A$ in explicit form for linear PDEs allows for the direct application of model-based linear controllers.

Similar to nonlinear PDEs, due to the numerical approximation of the spatial derivatives, one should choose $n_s$ to be sufficiently large to ensure the accuracy of the state-space model \eqref{eqn:linear_discrete}, with $n_s$ greater than about 50 sufficient in most cases. Due to the analytical temporal integration of the dynamics, there is no internal integration time step $dt$ to select. As in the nonlinear case, the sampling time $\Delta t$ may be chosen as large as desired. 

Since the state-space model \eqref{eqn:linear_discrete} is derived from a PDE, the eigenvalues and eigenvectors of the $A$ matrix can be analyzed explicitly. This method allows for a clearer understanding of the impact of the PDE's physical parameters on the system dynamics, such as open-loop stability. By adjusting these parameters, users can tailor the system dynamics to better assess their algorithms. We demonstrate this process with the convection-diffusion-reaction equation in Section \ref{sec:pde_eigens}, illustrating the relationships between its physical parameters and open-loop system dynamics.

\vspace{0.5em}
\noindent\textbf{Observation process. }
For all PDEs listed in Table \ref{table:pde}, we place $n_y$ sensors uniformly throughout the domain $\Omega$, where each sensor measures the unscaled value of the state at its location, perturbed by additive zero-mean Gaussian white noise. That is, the observation $y_k$ at time $k$ is computed by $y_k = Cs_k + v_k$, where $C \in \RR^{n_y \times n_s}$ is structured with a single $1$ per row and zeros elsewhere, and $v_k \sim \cN(0, \Sigma_v)$. Both $n_y$ and $\Sigma_v$ are user-configurable parameters. 

\vspace{0.5em}
\noindent\textbf{PDE control objectives. }For all PDEs listed in Table \ref{table:pde}, we define a control task whose primary objective is to steer the system's dynamics toward a user-defined target state $s_{ref} \in \RR^{n_s}$. The reward function is formulated as the negative sum of the LQ stage cost
\begin{align*}
	\cJ(a_k) = - \EE \left\{\sum_{k=0}^{K-1} \Big[(s_k - s_{ref})^{\top}Q(s_k-s_{ref}) + a_k^{\top}Ra_k\Big]\right\},
\end{align*} 
where $Q$ and $R$ are positive-definite weighting matrices that balance tracking performance and control effort. When the target state is the zero vector, the tracking problem reduces to the LQ regulation problem. 

\subsubsection{Convection-Diffusion-Reaction Equation}\label{sec:cdr}
\begin{figure}[h]
	\centering
	\includegraphics[width = 0.99\linewidth]{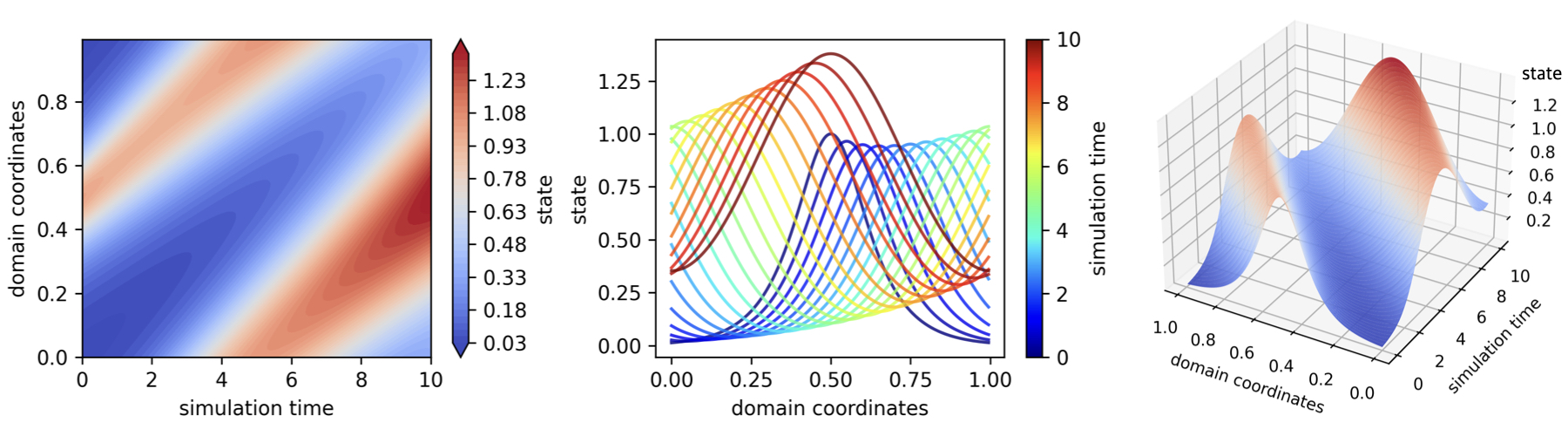}
	\caption{The uncontrolled solution to the CDR equation in a domain of length $L=1$ with parameters, $c=0.01$, $\nu = 0.002$, and $r = 0.1$. The initial condition is $u(x, t=0) = \mathrm{sech}(10x - 5)$. \textit{Left}: Contour plot that shows the value of the state variable over the total simulation time ($x$-axis) and across the spatial domain ($y$-axis). \textit{Middle}: Lines representing the state variable at fixed times. The $x$- and $y$-axes represent spatial coordinates and values of the state variable, respectively. The color of the lines corresponds to different time stamps within the total simulation time. \textit{Right}: 3D surface plot showing the value of the state variable ($z$-axis) over time ($y$-axis) and across the spatial domain ($x$-axis). }
	\label{fig:cdr}
\end{figure}

The convection-diffusion-reaction (CDR) equation models the transfer of particles, energy, or other physical quantities within a system due to convection, diffusion, and reaction processes. The temporal dynamics of the continuous concentration function $u(x, t)$ in one spatial dimension is given~by
\begin{align}
\frac{\partial u}{\partial t} + c \, \frac{\partial u}{\partial x} - \nu \, \frac{\partial^2 u}{\partial x^2} - r \, u =  a(x, t), \label{eqn:cdr}
\end{align}
where $c$ is the convection velocity, $\nu > 0$ is the diffusivity constant, $r$ is the reaction constant, and $a(x, t)$ is a source term defined in \eqref{eqn:distributed_control} that models the addition of energy to the system or other external influences that affect the PDE dynamics. The scalar physical parameters of the CDR equation characterize the strength of convection, diffusion, and reaction processes. When $c = r = 0$, the CDR equation \eqref{eqn:cdr} reduces to the heat equation. The CDR equation with $r=0$ has been used to validate the global convergence of RL algorithms in Kalman filtering \citep{zhang2023global}. We visualize the uncontrolled solution of the CDR equation for a specific choice of parameters and initial condition in Figure \ref{fig:cdr}.

After discretizing space and time, the CDR equation can be approximated by the linear state-space model \eqref{eqn:linear_discrete} with
\begin{align}\label{eqn:cdr_A_matrix}
	A = \frac{1}{n_s} W^\dagger \diag (e^{(-i c k_x - \nu k_x^2 + r) \Delta t}) W, \quad B_2 = \left[\arraycolsep=2pt\def\arraystretch{0.1}\begin{array}{c|c|c|c}
    \Phi_{0}    & \Phi_{1}    & \cdots & \Phi_{n_a - 1}
  \end{array}\right] \in \RR^{n_s \times n_a} 
\end{align}
where $i$ is the imaginary unit, $k_x = \frac{2 \pi}{L} [0, \dots, \frac{n_s}{2}-1, 0, -\frac{n_s}{2}+1, \dots, -1] \in \RR^{n_s}$ is the vector of spatial wavenumbers, $\Phi_j \in \RR^{n_s}$ are the forcing support functions evaluated at $n_s$ equally-spaced points in $\Omega$, and $W \in \RR^{n_s \times n_s}$ is the discrete Fourier transform (DFT) matrix with entries defined by \citep{rao2018transform}
\begin{align}\label{eqn:DFT_matrix}
	\forall p, q, \quad W_{pq} = e^{-\frac{2 \pi i (p-1)(q-1)}{n_s}}.
\end{align}
The scaled conjugate transpose of $W$, denoted by $W^\dagger/n_s$, is the inverse DFT matrix. In Section \ref{sec:pde_eigens}, we present the analytical eigenvalues and eigenvectors of the $A$ matrix in \eqref{eqn:cdr_A_matrix}, derived as functions of the physical parameters of the CDR equation.

\subsubsection{Wave Equation}
\begin{figure}[h]
	\centering
	\includegraphics[width = 0.99\linewidth]{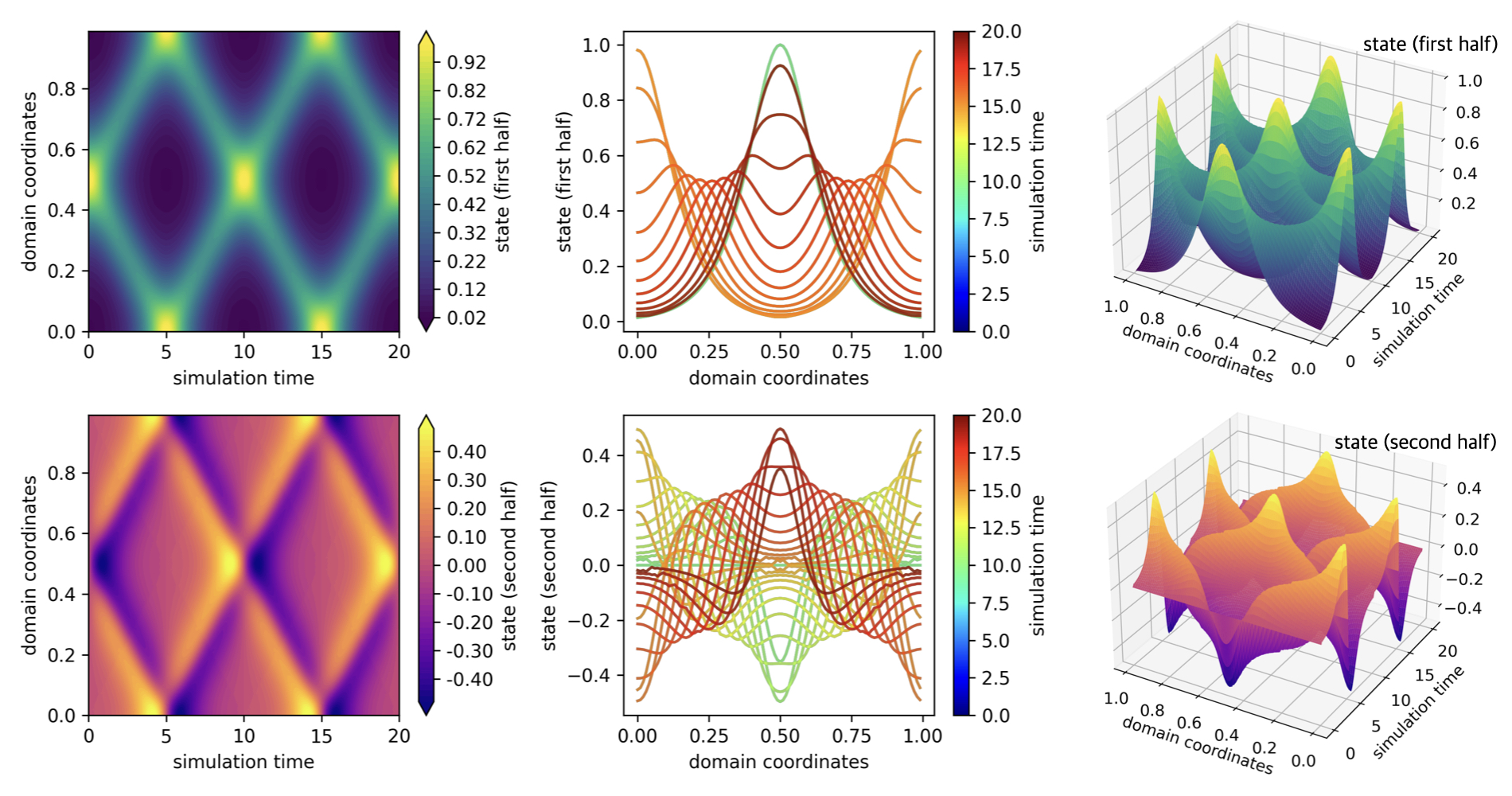}
	\caption{The uncontrolled solution to the wave equation in a domain of length $L=1$ with parameter $c=0.1$. The initial conditions are $u(x, t=0) = \mathrm{sech}(10x-5)$ and $\psi(x, t=0) = 0$. The figure convention is consistent with that of Figure \ref{fig:cdr}.}
	\label{fig:wave}
\end{figure}

The wave equation is a fundamental linear PDE in physics and engineering, describing the propagation of various types of waves through a homogeneous medium. The temporal dynamics of the perturbed scalar quantity $u(x, t)$ propagating as a wave through one-dimensional space is given by
\begin{align}\label{eqn:wave}
\frac{\partial^2 u}{\partial t^2} - c^2 \, \frac{\partial^2 u}{\partial x^2} = a(x, t),
\end{align}
where $c$ is a constant representing the wave's speed in the medium, and $a(x, t)$ is a source term defined in \eqref{eqn:distributed_control} that models the effect of a force or other external influences acting on the system. We visualize the uncontrolled solution of the wave equation for a specific choice of parameters and initial condition in Figure \ref{fig:wave}.

We solve the wave equation by first transforming \eqref{eqn:wave} into a coupled system of two PDEs with first-order time derivatives. Specifically, we introduce a second continuous field $\psi(x, t):\Omega \times \mathbb{R}^+ \rightarrow \mathbb{R}$ representing the rate at which the scalar quantity $u(x, t)$ is changing locally. Then, we can write \eqref{eqn:wave} in the equivalent form
\begin{subequations}\label{eqn:wave_first_order}
\begin{align}
	\frac{\partial u}{\partial t} - \psi &= 0, \\
	\frac{\partial \psi}{\partial t} - c^2 \, \frac{\partial^2 u}{\partial x^2} &= a(x, t).
\end{align}
\end{subequations}
We next discretize space and time by introducing a state vector $s_k$ that concatenates the values of both $u$ and $\psi$, each sampled at $n_s' = n_s/2$ equally-spaced points in $\Omega$ so that $s_k$ contains $n_s$ components. The wave equation in the form \eqref{eqn:wave_first_order} can then be approximated by the state-space model \eqref{eqn:linear_discrete} with
\begin{align}\label{eqn:wave_A_matrix}
	A &= \frac{1}{n_s'} \cdot \left[ 
\begin{array}{c|c}
W & 0_{n_s' \times n_s'} \\ \hline
0_{n_s' \times n_s'} & W
\end{array} 
\right]^\dagger
	\exp \left( \left[ 
\begin{array}{c|c}
0_{n_s' \times n_s'} & I_{n_s' \times n_s'} \\ \hline
\Lambda & 0_{n_s' \times n_s'}
\end{array} 
\right] \Delta t \right)
	\left[ 
\begin{array}{c|c}
W & 0_{n_s' \times n_s'} \\ \hline
0_{n_s' \times n_s'} & W
\end{array} 
\right] \\
B_2 &= \left[\begin{array}{c|c|c|c}
 0_{n_s'} & 0_{n_s'} & \cdots & 0_{n_s'} \\ \hline
    \Phi_{0}'    & \Phi_{1}'    & \cdots & \Phi_{n_a - 1}'
  \end{array}\right] \in \RR^{n_s \times n_a},  \nonumber
\end{align}
where $\Lambda = \diag(-c^2 k_x^2) \in \RR^{n_s' \times n_s'}$ with $k_x = \frac{2 \pi}{L} [0, \dots, \frac{n_s'}{2}-1, 0, -\frac{n_s'}{2}+1, \dots, -1] \in \RR^{n_s'}$ the vector of spatial wavenumbers,  $\Phi_j' \in \RR^{n_s'}$ are the forcing support functions evaluated at $n_s'$ equally-spaced points in $\Omega$, and $W \in \RR^{n_s' \times n_s'}$ is the DFT matrix \eqref{eqn:DFT_matrix} but with $n_s$ replaced by $n_s'$.

\subsubsection{Schrödinger Equation}\label{sec:schrodinger}
\begin{figure}[h]
	\centering
	\includegraphics[width = 0.99\linewidth]{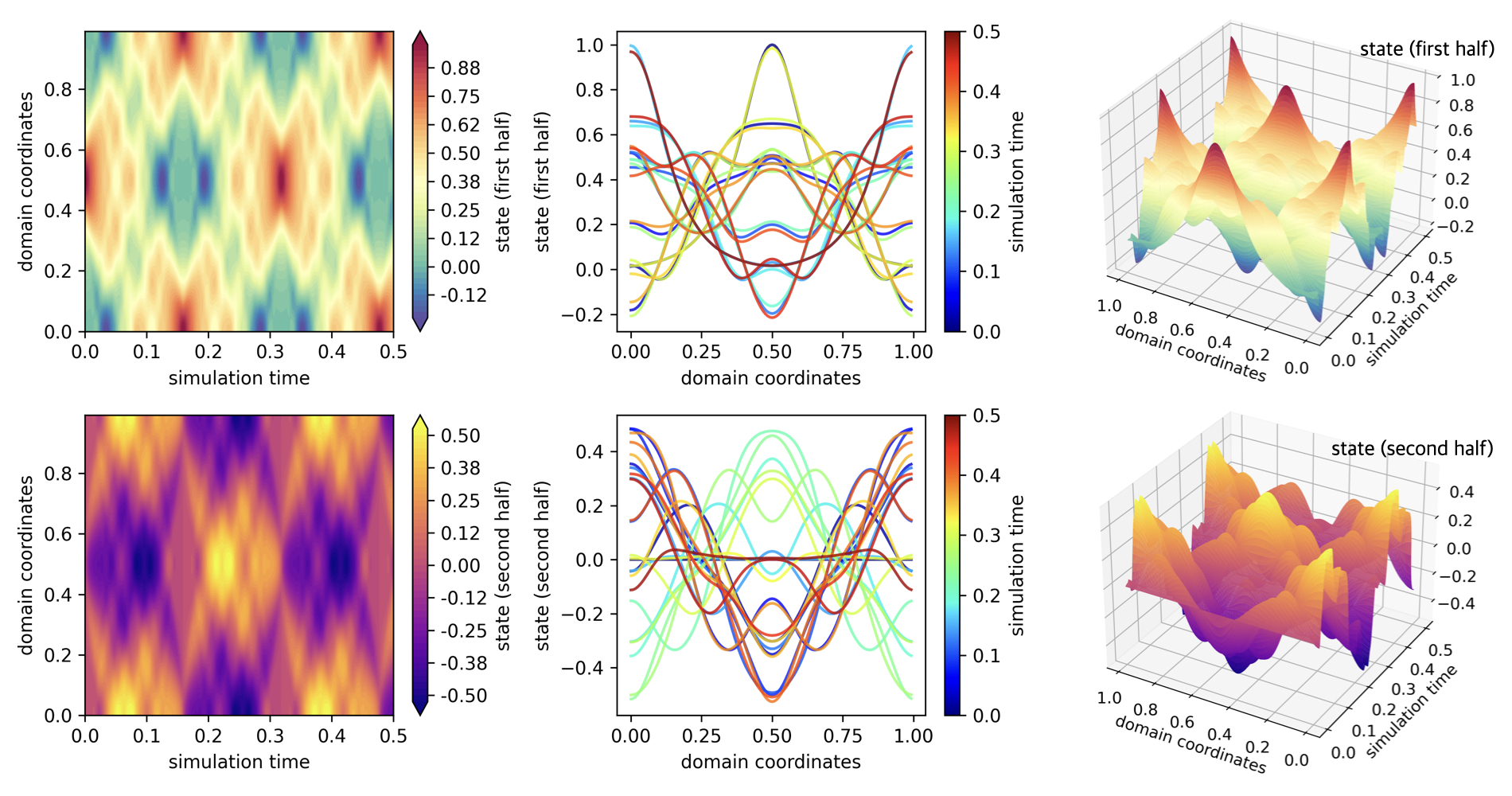}
	\caption{The uncontrolled solution to the Schrödinger equation in a domain of length $L=1$ with parameters $\hbar = 1.0$, $m=1.0$, and $V=0.0$. The initial conditions are $\xi(x, t=0)=\mathrm{sech}(10x-5)$ and $\eta(x, t=0) = 0$. The figure convention is consistent with that of Figure \ref{fig:cdr}.}
	\label{fig:schrodinger}
\end{figure}

The Schrödinger equation is fundamental in quantum mechanics, describing how the quantum state of an isolated quantum-mechanical system, a complex-valued wave function, changes over time. For a single non-relativistic particle in a constant potential, the Schrödinger equation for the wave function $u(x, t)$ is given by
\begin{align}\label{eqn:schrodinger}
i\hbar \frac{\partial u}{\partial t} +\frac{\hbar^2}{2m} \frac{\partial^2 u}{\partial x^2} - V \, u = \hbar \, a(x, t),
\end{align}
where $i$ is the imaginary unit, $\hbar$ is the Planck constant, $V$ is the real potential constant, and $a(x, t)$ is a real-valued source term defined in \eqref{eqn:distributed_control} that models a force acting on the system or other external influences that affect the PDE dynamics. We visualize the uncontrolled solution of the Schrödinger equation for a specific choice of parameters and initial condition in Figure \ref{fig:schrodinger}.

We solve the Schrödinger equation by first transforming \eqref{eqn:schrodinger} into a coupled system of two PDEs with first-order time derivatives, similar to the approach adopted for the wave equation. Specifically, we introduce two continuous fields $\xi(x, t):\Omega \times \mathbb{R}^+ \rightarrow \mathbb{R}$ and $\eta(x, t):\Omega \times \mathbb{R}^+ \rightarrow \mathbb{R}$ that represent the real and imaginary parts of the complex-valued scalar quantity $u(x, t)$, respectively. Then, we rewrite \eqref{eqn:wave} in the equivalent form
\begin{subequations}
\begin{align}
	\frac{\partial \xi}{\partial t} + \frac{\hbar}{2m} \frac{\partial^2 \eta}{\partial x^2} - \frac{V}{\hbar} \eta &= 0, \\
	\frac{\partial \eta}{\partial t} - \frac{\hbar}{2m} \frac{\partial^2 \xi}{\partial x^2} + \frac{V}{\hbar} \xi &= a(x, t),
\end{align}
\end{subequations}
We now discretize space and time by introducing a state vector $s_k$ that concatenates the values of both $\xi$ and $\eta$, each sampled at $n_s' = n_s/2$ equally-spaced points in $\Omega$ so that $s_k$ contains $n_s$ components. The Schrödinger equation in the form \eqref{eqn:wave_first_order} can then be approximated by the state-space model \eqref{eqn:linear_discrete} with
\begin{align}\label{eqn:schrodinger_A_matrix}
	A &= \frac{1}{n_s'} \left[ 
\begin{array}{c|c}
W & 0_{n_s' \times n_s'} \\ \hline
0_{n_s' \times n_s'} & W
\end{array} 
\right]^\dagger
	\exp \left( \left[ 
\begin{array}{c|c}
0_{n_s' \times n_s'} & \Lambda \\ \hline
-\Lambda & 0_{n_s' \times n_s'}
\end{array} 
\right] \Delta t \right)
	\left[ 
\begin{array}{c|c}
W & 0_{n_s' \times n_s'} \\ \hline
0_{n_s' \times n_s'} & W
\end{array} 
\right], \\
B_2 &= \left[\begin{array}{c|c|c|c}
 0_{n_s'} & 0_{n_s'} & \cdots & 0_{n_s'} \\ \hline
    \Phi_{0}' & \Phi_{1}' & \cdots & \Phi_{n_a - 1}'
  \end{array}\right] \in \RR^{n_s \times n_a},  \nonumber
\end{align}
where $\Lambda = \diag(\hbar/(2m) k_x^2 + V/\hbar) \in \RR^{n_s' \times n_s'}$ with $k_x = \frac{2 \pi}{L} [0, \dots, \frac{n_s'}{2}-1, 0, -\frac{n_s'}{2}+1, \dots, -1] \in \RR^{n_s'}$ the vector of spatial wavenumbers, $\Phi_j' \in \RR^{n_s'}$ are the forcing support functions evaluated at $n_s'$ equally-spaced points in $\Omega$, and $W \in \RR^{n_s' \times n_s'}$ is the DFT matrix defined in \eqref{eqn:DFT_matrix} but with $n_s$ therein replaced by $n_s'$.

\subsubsection{Burgers' Equation}
\begin{figure}[h]
	\centering
	\includegraphics[width = 0.99\linewidth]{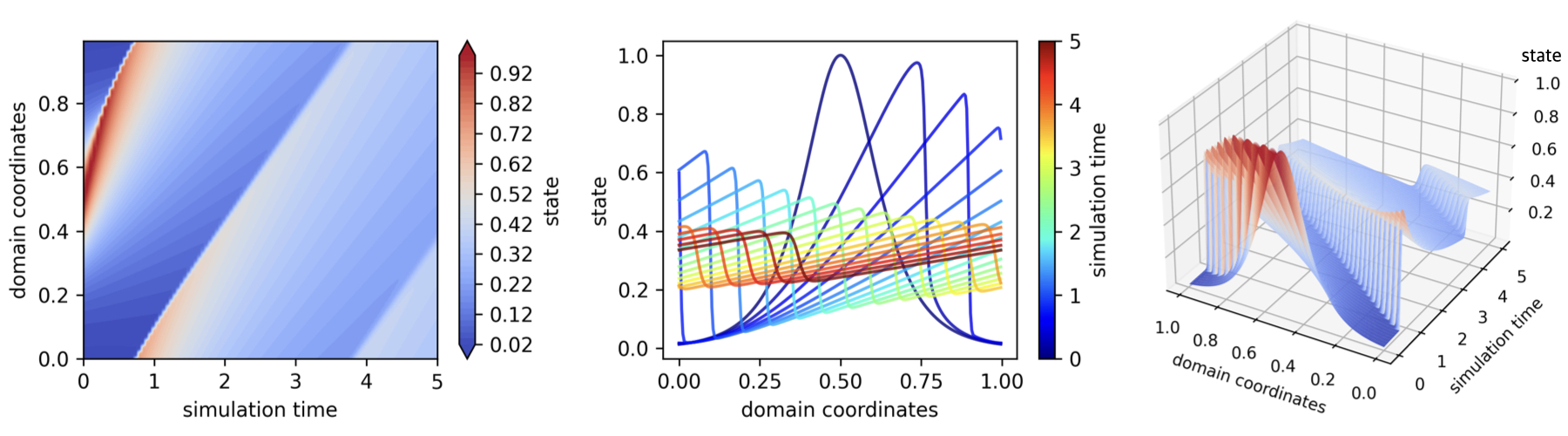}
	\caption{The uncontrolled solution to Burgers' equation in a domain of length $L=1$ with parameter $\nu = 0.001$. The initial condition is $u(x, t=0) = \mathrm{sech}(10x - 5)$. The figure convention is consistent with that of Figure \ref{fig:cdr}.}
	\label{fig:burgers}
\end{figure}

Burgers' equation is a simplified version of nonlinear PDEs arising in fluid dynamics and captures key features of water waves and gas dynamics such as shock formation. The temporal dynamics of the velocity $u(x, t)$ is
\begin{align}
\frac{\partial u}{\partial t} + u\frac{\partial u}{\partial x} - \nu \frac{\partial^2 u}{\partial x^2} = a(x, t),\label{eqn:burgers}
\end{align}
where $\nu > 0$ is the diffusivity (or viscosity) parameter and $a(x, t)$ is a source term defined in \eqref{eqn:distributed_control} that models a force acting on the system or other external influences that affect the PDE dynamics. At the inviscid limit of $\nu = 0$, Burgers' equation predicts discontinuous shocks; at low $\nu$ values, Burgers' equation exhibits shock-like behavior but remains smooth; and with high values of $\nu$, Burgers' equation mirrors the dissipative nature of the heat equation. The behavior of the uncontrolled solution for a specific choice of parameters and initial condition is shown in Figure \ref{fig:burgers}.

\subsubsection{Kuramoto-Sivashinsky Equation}
\begin{figure}[h]
	\centering
	\includegraphics[width = 0.99\linewidth]{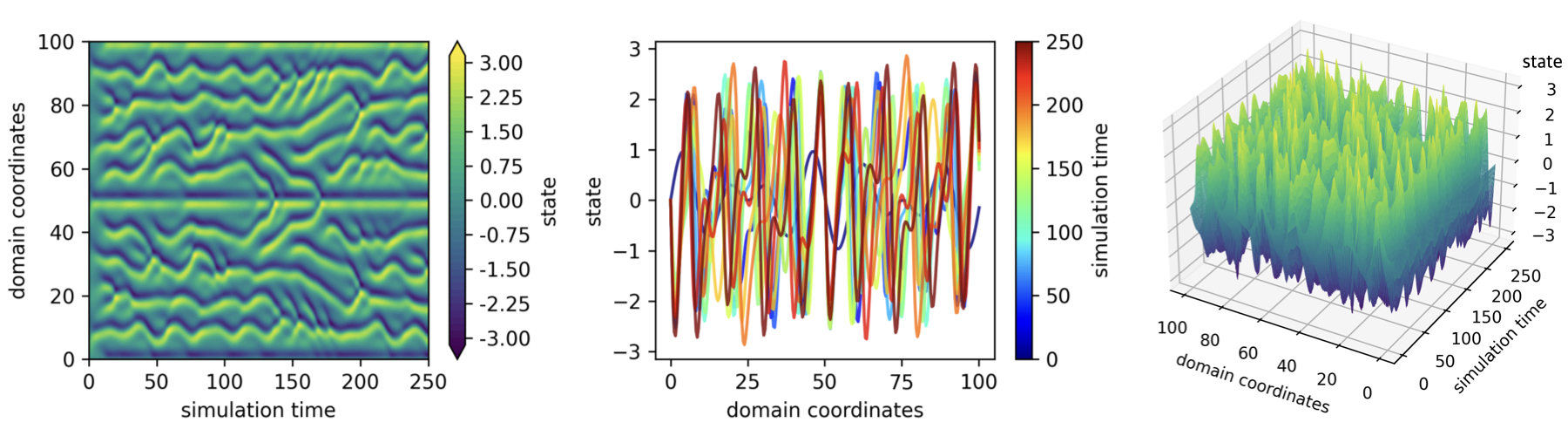}
	\caption{The uncontrolled solution to the KS equation in a domain of length $L = 32 \pi$. The initial condition is $u(x, t=0) = \frac{1}{16} x \, \sin(\frac{3}{8}x)$ . The figure convention is consistent with that of Figure \ref{fig:cdr}.}
	\label{fig:ks}
\end{figure}

The Kuramoto-Sivashinsky (KS) equation is a nonlinear PDE applied to studying pattern formation and instability in fluid dynamics, combustion, and plasma physics. The temporal dynamics of $u(x, t)$ in one spatial dimension is provided by
\begin{align}\label{eqn:ks}
\frac{\partial u}{\partial t} + u\frac{\partial u}{\partial x} + \frac{\partial^2 u}{\partial x^2} + \frac{\partial^4 u}{\partial x^4} = a(x, t),
\end{align}
where $a(x, t)$ is a source term defined in \eqref{eqn:distributed_control} that models a force acting on the system or other external influences that affect the dynamics. The nonlinear convection term, the second-order diffusion term, and the fourth-order dispersion term interact to produce complex spatial patterns and temporal chaos when the domain length $L$ is large enough \citep{cvitanovic2010state}. Figure \ref{fig:ks} displays the behavior of the uncontrolled solution of the KS equation for a specific initial condition and $L = 32 \pi$, well into the chaotic regime. Due to the chaotic nature of the dynamics, the specific choice of initial condition has negligible influence on the qualitative properties of the solution.

\subsubsection{Fisher Equation}
\begin{figure}[h]
	\centering
	\includegraphics[width = 0.99\linewidth]{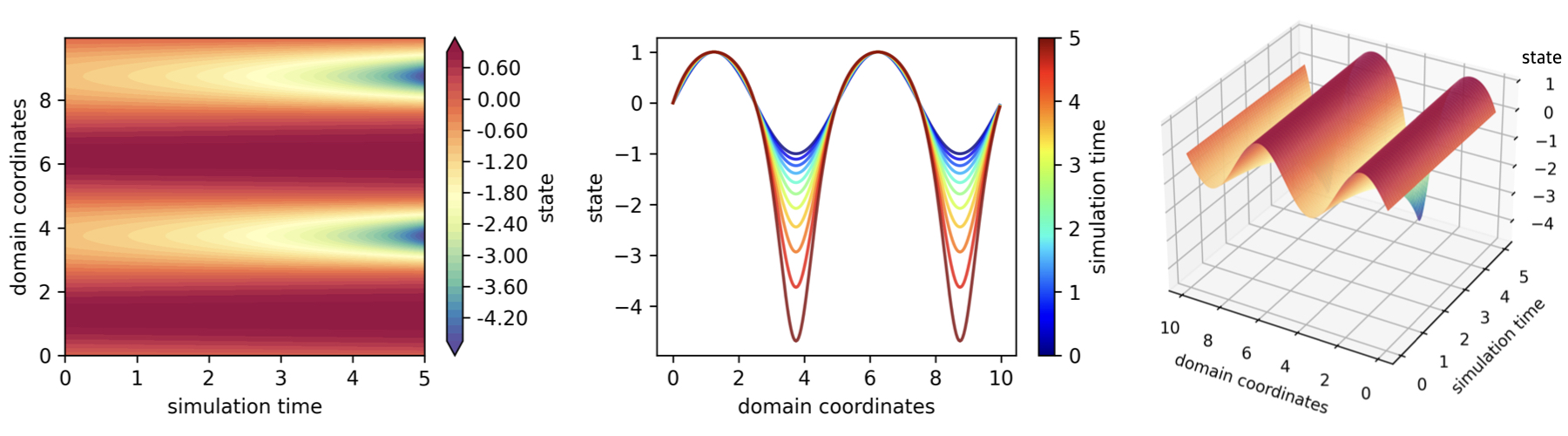}
	\caption{The uncontrolled solution to the Fisher equation in a domain of length $L = 10$ with parameters $\nu = 0.0001$ and $r = 0.1$. The initial condition is $u(x, t=0) = \sin(\frac{2 \pi x}{5})$. The figure convention is consistent with that of Figure \ref{fig:cdr}. }
	\label{fig:fisher}
\end{figure}

The Fisher equation is a nonlinear PDE employed in biology, ecology, and epidemiology to model gene propagation, invasions, and population dynamics. The temporal dynamics of $u(x, t)$ in one spatial dimension is described by
\begin{align}\label{eqn:fisher}
	\frac{\partial u}{\partial t} - \nu \frac{\partial^2 u}{\partial x^2} - r \cdot u(1-u) = a(x, t),
\end{align}
where $\nu > 0$ is the diffusivity constant, $r$ is the reaction constant, and $a(x, t)$ is a source term defined in \eqref{eqn:distributed_control} that models external influences affecting the PDE dynamics. The term $r\cdot u(1-u)$ captures population expansion limited by carrying capacity, with $r$ as the intrinsic growth rate. The uncontrolled solution of the Fisher equation for a specific choice of parameters and initial condition is depicted in Figure \ref{fig:fisher}.

\subsubsection{Allen-Cahn Equation}
\begin{figure}[h]
	\centering
	\includegraphics[width = 0.99\linewidth]{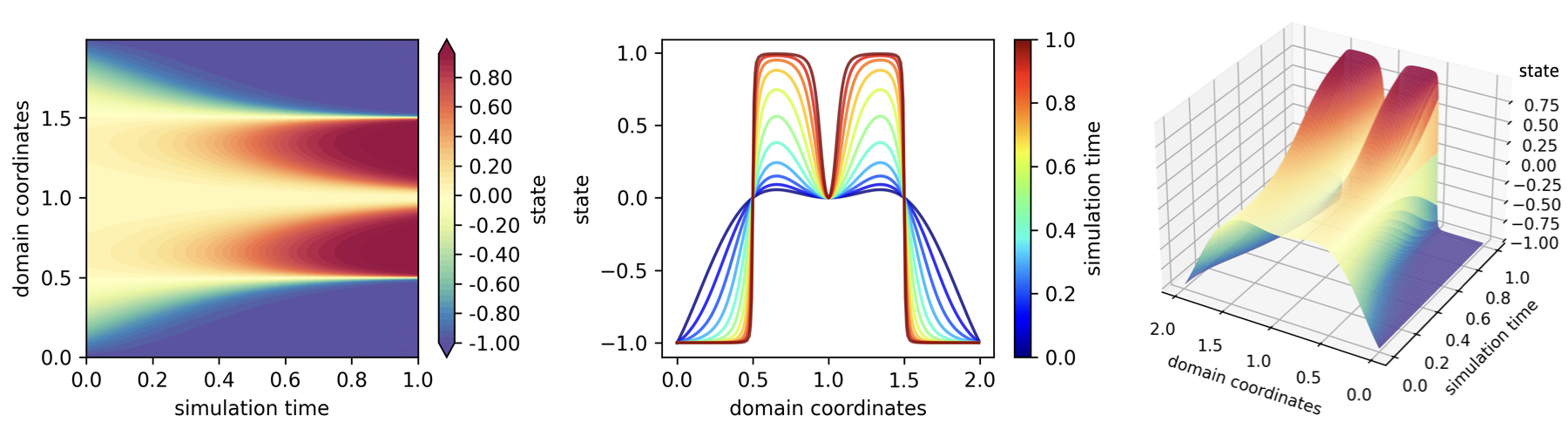}
	\caption{The uncontrolled solution to the Allen-Cahn equation in a domain of length $L = 2$ with parameters $\nu = 0.0001$ and $V = 5.0$. The initial condition is $u(x, t=0) = (x-1)^2\cdot \cos(\pi(x-1))$. The figure convention is consistent with that of Figure \ref{fig:cdr}.}
	\label{fig:allen_cahn}
\end{figure}

The Allen-Cahn equation is a nonlinear PDE modeling phase separation in binary alloy systems in materials science. The temporal dynamics of $u(x, t)$ in one spatial dimension, with $u = \pm 1$ indicating the presence of one phase or the other, is given by
\begin{align}\label{eqn:allen_cahn}
\frac{\partial u}{\partial t} - \nu^2 \frac{\partial^2 u}{\partial x^2} + V(u^3 - u) = a(x, t),
\end{align}
where $\nu > 0$ is the diffusivity constant, $V$ is the potential constant, and $a(x, t)$ is a source term defined in \eqref{eqn:distributed_control} that models external influences affecting the PDE dynamics. We visualize the uncontrolled solution of the Allen-Cahn equation for a specific choice of parameters and initial condition in Figure \ref{fig:allen_cahn}.

\subsubsection{Korteweg-de Vries Equation}
\begin{figure}[h]
	\centering
	\includegraphics[width = 0.99\linewidth]{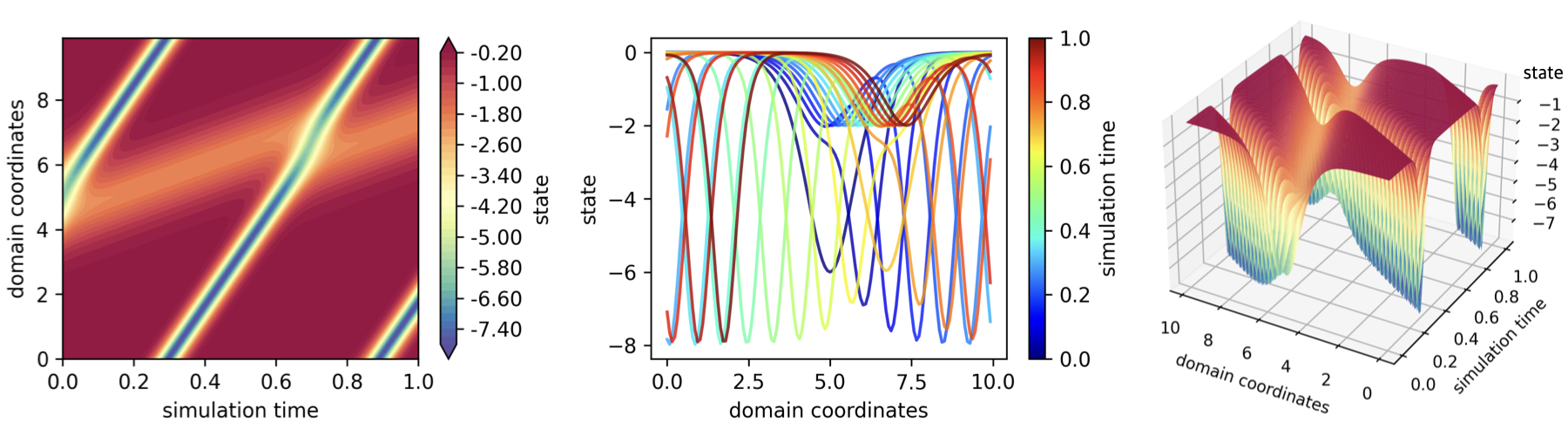}
	\caption{The uncontrolled solution to the KdV equation in a domain of length $L = 10$. The initial condition is $u(x, t=0) = \frac{-12(3+\cosh(20-4x)+4\cosh(10-2x))}{(\cosh(15-3x) + 3\cosh(5-x))^2}$. The figure convention is consistent with that of Figure \ref{fig:cdr}.}
	\label{fig:kdv}
\end{figure}

The Korteweg-de Vries (KdV) equation is a nonlinear PDE pivotal in understanding nonlinear wave dynamics, modeling solitary wave propagation across shallow water surfaces, with applications extending to plasma physics, nonlinear optics, and quantum mechanics. The temporal dynamics of $u(x, t)$ with an additional source term $a(x, t)$ that models external influences is given by
\begin{align}\label{eqn:kdv}
\frac{\partial u}{\partial t} + \frac{\partial^3 u}{\partial x^3} - 6u\frac{\partial u}{\partial x} = a(x, t).
\end{align}
We visualize the uncontrolled solution of the KdV equation for a specific choice of parameters and an initial condition that leads to two solitons propagating at different speeds in Figure \ref{fig:kdv}.

\subsubsection{Cahn-Hilliard Equation}
\begin{figure}[h]
	\centering
	\includegraphics[width = 0.99\linewidth]{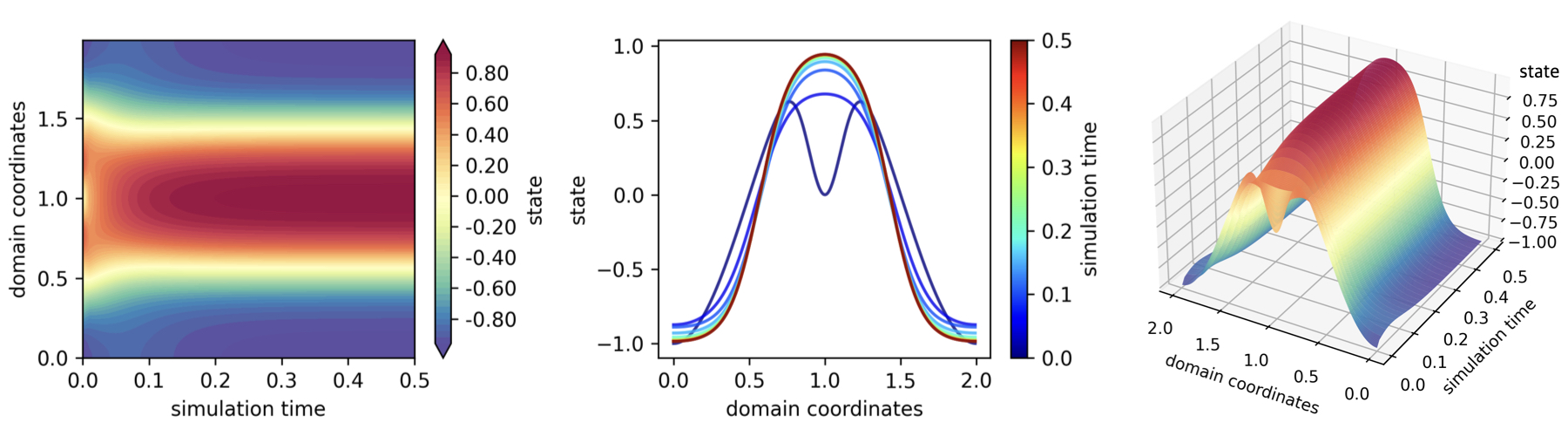}
	\caption{The uncontrolled solution to the Cahn-Hilliard equation in a domain of length $L=2$ with parameters $\nu=1.0$ and $\Gamma=0.02$. The initial condition is $u(x, t=0)=\cos(\pi(x-1)) - \exp(-4(\pi(x-1))^2)$. The figure convention is consistent with that of Figure \ref{fig:cdr}.}
	\label{fig:cahn_hilliard}
\end{figure}

The Cahn-Hilliard equation is a nonlinear PDE modeling phase separation in alloys and polymers in materials science. The temporal dynamics of $u(x, t)$ in one spatial dimension, with $u = \pm 1$ indicating the presence of one phase or the other, is described by
\begin{align}\label{eqn:cahn_hilliard}
\frac{\partial u}{\partial t} - \nu \frac{\partial^2}{\partial x^2} (u^3 - u - \Gamma \frac{\partial^2u}{\partial x^2}) = a(x, t),
\end{align}
where $\nu > 0$ is the diffusivity constant, $\Gamma$ is the constant surface tensor coefficient, and $a(x, t)$ is a source term defined in \eqref{eqn:distributed_control} that models external influences affecting the PDE dynamics. We visualize the uncontrolled solution of the Cahn-Hilliard equation for a specific choice of parameters and initial condition in Figure \ref{fig:cahn_hilliard}.

\subsubsection{Ginzburg-Landau Equation}\label{sec:ginzburg_landau}
\begin{figure}[h]
	\centering
	\includegraphics[width = 0.99\linewidth]{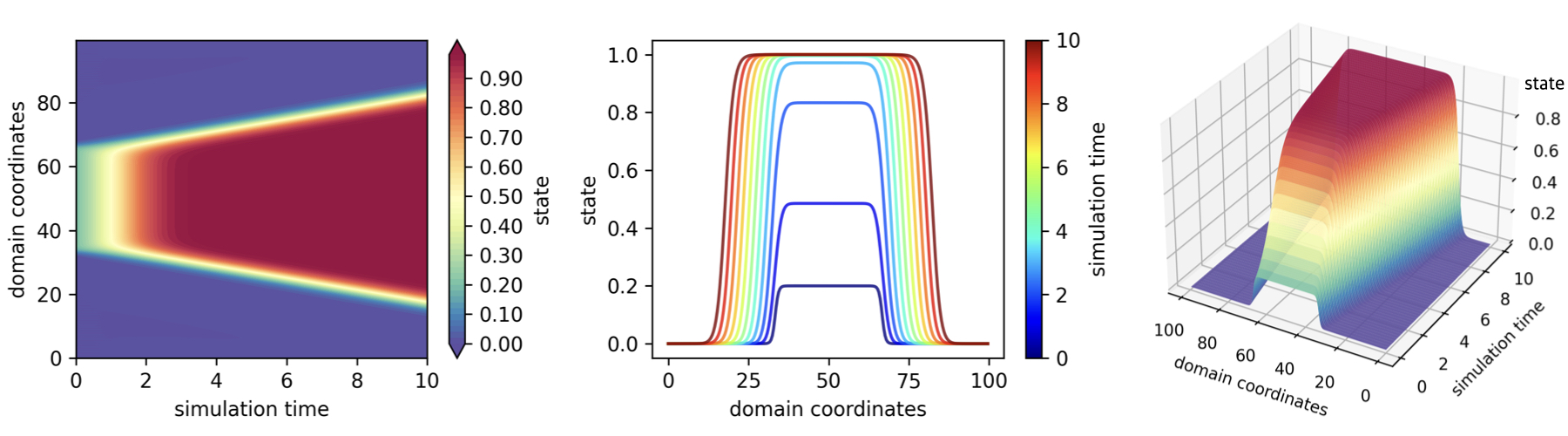}
	\caption{The uncontrolled solution to the Ginzburg-Landau equation in a domain of length $L=100$. The initial condition is $u(x, t=0) = \frac{1}{10}(\tanh(x-\frac{100}{3}) - \tanh(x-\frac{200}{3}))$. The figure convention is consistent with that of Figure \ref{fig:cdr}.}
	\label{fig:ginzburg_landau}
\end{figure}

The Ginzburg-Landau equation is a nonlinear PDE describing the evolution of disturbances near the onset of instability in various physical systems. The temporal dynamics of the amplitude $u(x, t)$ of a disturbance in one spatial dimension is governed by
\begin{align}\label{eqn:ginzburg_landau}
\frac{\partial u}{\partial t} - u + |u|^2 u - \frac{\partial^2 u}{\partial x^2} = a(x, t),
\end{align}
where $a(x, t)$ is a source term defined in \eqref{eqn:distributed_control} that models external influences affecting the PDE dynamics. We visualize the uncontrolled solution of the Ginzburg-Landau equation for a specific choice of parameters and initial condition in Figure \ref{fig:ginzburg_landau}. 

\section{Examples of Using \texttt{controlgym}}
This section provides several examples for using \texttt{controlgym}. In particular, we include a detailed analysis of how users can design the open-loop dynamics of linear PDEs by selecting physical parameters in Section \ref{sec:pde_eigens}. Sections \ref{sec:model_based_control} and \ref{sec:model_free_learning} provide code examples of applying a model-based controller (as baseline) and an RL-based controller to a helicopter environment, respectively. Lastly, we demonstrate in Section \ref{sec:zero_control} how to obtain the uncontrolled PDE trajectories displayed in Figures \ref{fig:cdr}-\ref{fig:ginzburg_landau} using a code example.

\subsection{Analysis of the Open-Loop Dynamics of Linear PDEs}\label{sec:pde_eigens}

Evaluating learning algorithms effectively involves designing and tuning the open-loop dynamics of control environments, for adjusting the control difficulties. Precisely, in linear PDE environments, the open-loop dynamics is entirely specified by the spectral properties of matrix $A$ in \eqref{eqn:linear_discrete}. Using the CDR equation from Section \ref{sec:cdr} as a case study, we demonstrate the analytical derivation of the open-loop system dynamics and its connection to the physical parameters $c$, $\nu$, and $r$ in the corresponding PDE \eqref{eqn:cdr}. The methodology that we follow is applicable to any linear PDEs with constant physical parameters \citep{cross1993pattern,schmid2001stability}.

First, we rewrite the uncontrolled CDR equation \eqref{eqn:cdr} as
\begin{align}
\frac{\partial u}{\partial t} - \cL u = 0, \quad \cL = - c \, \frac{\partial}{\partial x} + \nu \, \frac{\partial^2}{\partial x^2} + r, \label{eqn:cdr_uncontrolled}
\end{align}
Eigenvalues $\omega$ and eigenfunctions $\alpha(x)$ of the linear differential operator $\cL$ are defined by the relation $\cL\alpha = \omega\alpha$. For a PDE with constant physical parameters in a periodic domain with length $L$, all eigenfunctions $\alpha(x)$ have the form of $\alpha(x) = e^{i k x}$, where the admissible wavenumbers $k$ are calculated by $k = p k_0$ with $p \in \mathbb{N}$ and $k_0 = \frac{2\pi}{L}$. The corresponding eigenvalues $\omega$ are obtained by applying $\cL$ to $\alpha$; that is,
\begin{align}
\mathcal{L} e^{i k x}  = \left( - c \, \frac{\partial}{\partial x} + \nu \, \frac{\partial^2}{\partial x^2} + r \right) e^{i k x} = (-ick - \nu k^2 + r) e^{i k x} = \omega e^{i k x}.
\end{align}

Hence, $\omega = -ick - \nu k^2 + r$ is the eigenvalue corresponding to the eigenfunction $e^{i k x}$ for any $k = p k_0$, $p \in \mathbb{N}$. Denoting the real and imaginary parts of $\omega$ as $\omega_r = - \nu k^2 + r$ and $\omega_i = -ck$, respectively, we can write general solutions to \eqref{eqn:cdr_uncontrolled} as
\begin{align}\label{eqn:cdr_A_eigen}
u(x,t) = \sum_k e^{\omega t} e^{ikx} = \sum_k e^{\omega_r t} e^{i(x+\omega_i t)}.
\end{align}

Equation \eqref{eqn:cdr_A_eigen} shows that the eigenfunction $e^{i k x}$ either grows or decays exponentially at rate $\omega_r$ and propagates spatially with a phase speed of $-\omega_i/k$. This allows us to employ the spectral properties of $\mathcal{L}$ to characterize the behavior of solutions to \eqref{eqn:cdr_uncontrolled}.

To determine the eigenvalues and eigenvectors of matrix $A$ in the discrete-time state-space model \eqref{eqn:linear_discrete}, we discretize space and time in \eqref{eqn:cdr_A_eigen}. Spatial discretization transforms the continuous eigenfunctions into eigenvectors defined by the values of $e^{ik x}$ at $n_s$ evenly distributed points within $\Omega$, where the admissible wavenumbers $k$ are the entries of the vector $k_x$ from Section \ref{sec:cdr}.  Temporally, $\omega$ is replaced with its discrete-time analogue $\lambda = e^{\omega \Delta t}$. Consequently, the eigenvalues of $A$ are $\lambda = e^{(-ick - \nu k^2 + r) \Delta t}$, where $k$ is an element of $k_x$.

\begin{figure}[h]
	\centering
	\includegraphics[width = 0.6\textwidth]{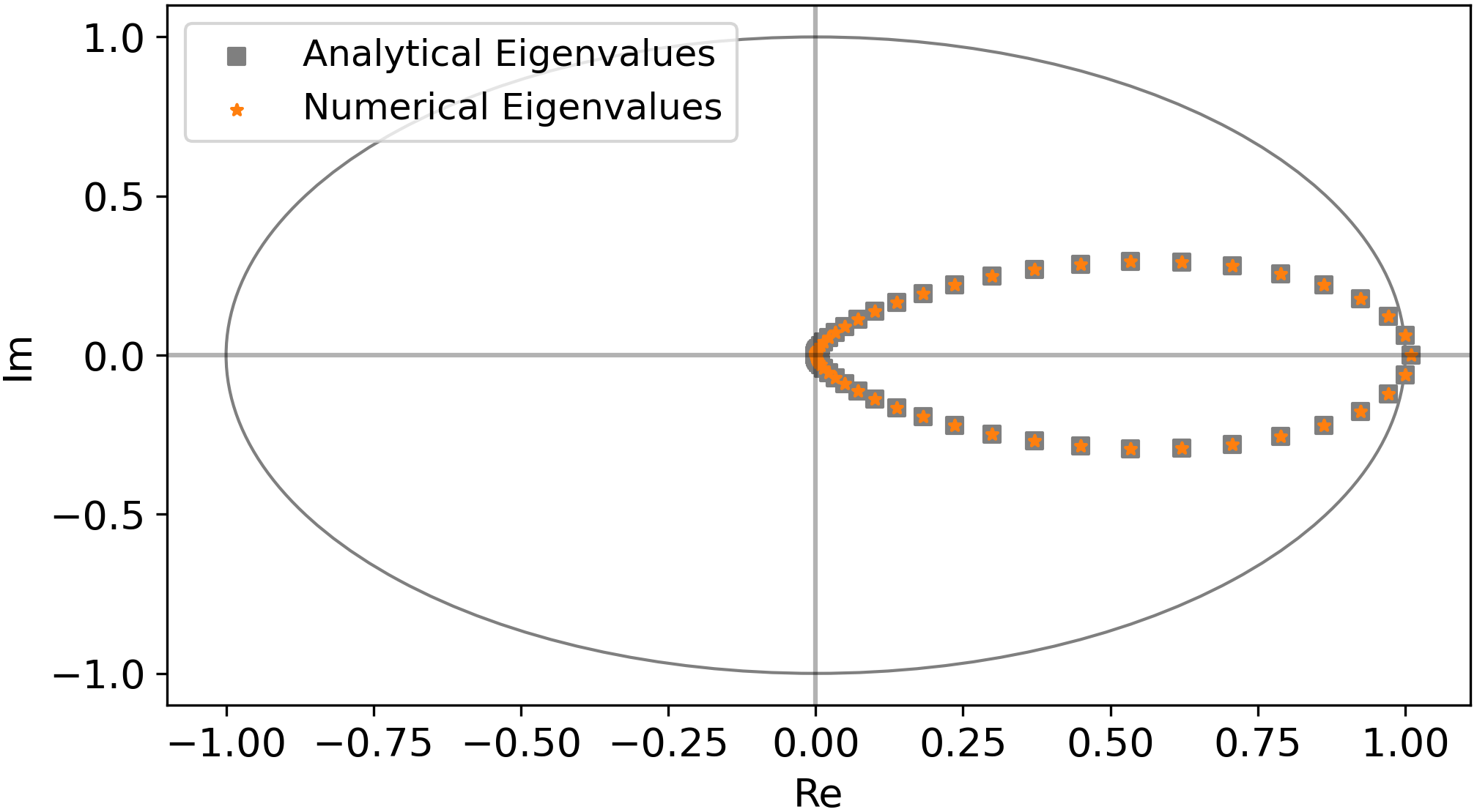}
	\caption{Analytical and numerical eigenvalues of the uncontrolled CDR equation}
	\label{fig:cdr_eigen}
\end{figure}

Figure \ref{fig:cdr_eigen} confirms the matching analytical and numerical eigenvalues of $A$. Our approach allows users to tune the system's open-loop stability by choosing $\nu$, $c$, and $r$. For example, the presence of one unstable eigenvalue outside the unit circle in Figure \ref{fig:cdr_eigen} explains the growth of the solution to the CDR equation seen in Figure \ref{fig:cdr}. By choosing a negative value for $r$, one can instead obtain an open-loop stable CDR equation environment. Lastly, we note that one can apply the same derivation process to the wave and Schrödinger equations.

\subsection{Model-Based Controllers as Baselines}\label{sec:model_based_control}
\begin{figure}[t]
\hspace{0.5em}
\begin{minipage}{0.56\textwidth}
\begin{lstlisting}[language=Python, style=mystyle]
import controlgym as gym

if __name__ == "__main__":
	env = gym.make("he1")
	lqg = gym.controllers.LQG(env)
	lqg.run()
	gym.save(lqg)
\end{lstlisting}
\end{minipage}\hspace{0.8em}
\begin{minipage}{0.4\textwidth}
	\includegraphics[height = 4.7cm]{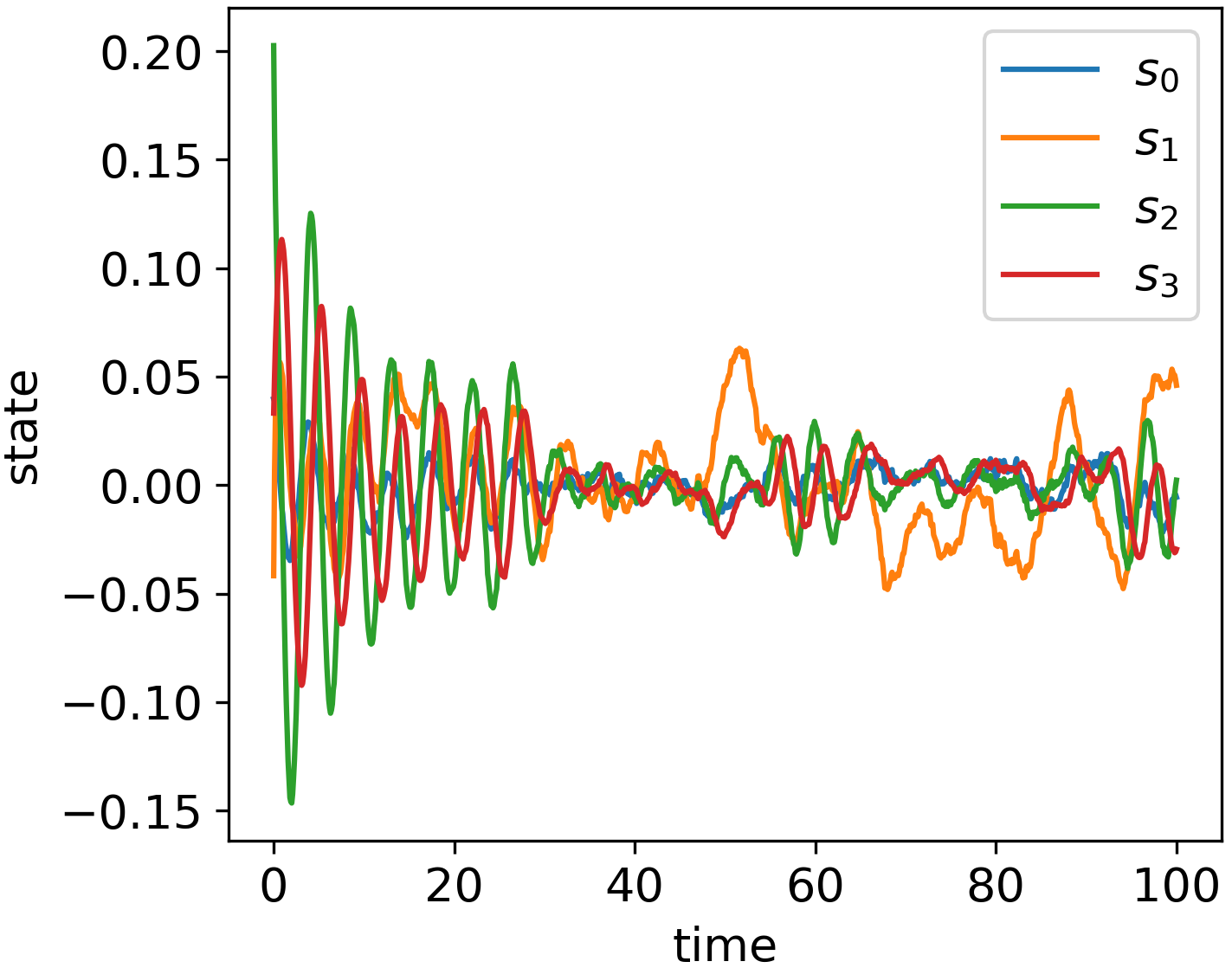}
\end{minipage}
\caption{LQG controller applied to environment ``he1". \textit{Left:} Python code snippet for setting up the environment and controller. \textit{Right:} Plot of the system states against time.}
\label{fig:he1_example}
\end{figure}
 In Figure \ref{fig:he1_example}, we show a code example of implementing the linear-quadratic-Gaussian (LQG) controller to the helicopter environment ``he1''. In this example, we set the sampling time $\Delta t$ to be $0.1$ and construct the environment using the function call \texttt{controlgym.make()}. Other environments could be set up similarly with environment IDs from Tables \ref{table:control}-\ref{table:pde} and optional keyword arguments. In addition to the LQG controller, we implement LQR and the state-feedback $H_2/H_{\infty}$ controllers in \texttt{controlgym}. For examples of applying baseline model-based controllers to linear PDE environments, we refer the readers to the example notebook file in our GitHub repository.  

\subsection{Model-Free RL Algorithms}\label{sec:model_free_learning}

\begin{figure}[t]
\hspace{0.5em}
\begin{minipage}{0.56\textwidth}
\begin{lstlisting}[language=Python, style=mystyle]
import controlgym as gym

if __name__ == "__main__":
	env = gym.make("he1")
	ppo = gym.controllers.PPO(env,  
	  actor_hidden_dim = 64, 
	  critic_hidden_dim = 64, lr = 1e-5)
	ppo.train(num_train_iter=100, 
	  num_episodes_per_iter=64, 
	  episode_length=100, sgd_epoch_num=4, 
	  mini_batch_size=5, cov_param=0.05)
	ppo.run()
	gym.save(ppo)
\end{lstlisting}
\end{minipage}\hspace{0.8em}
\begin{minipage}{0.4\textwidth}
	\includegraphics[height = 4.7cm]{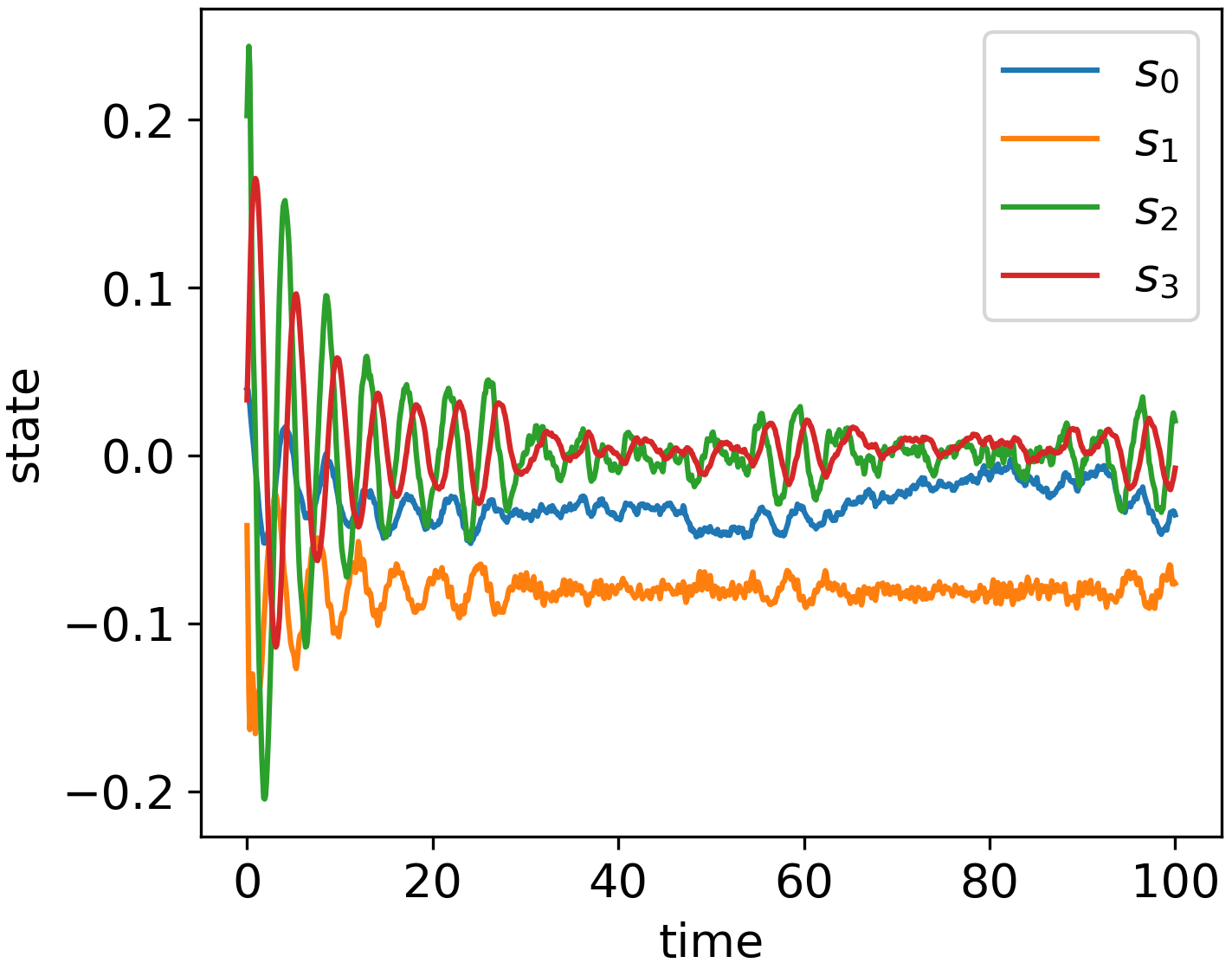}
\end{minipage}
\caption{PPO controller applied to the helicopter environment ``he1"}
\label{fig:ppo_he1_example}
\end{figure}

Other than the baseline model-based controllers, we also implement the proximal policy optimization (PPO) algorithm in \texttt{controlgym}. At the same time, all our environments support standard RL algorithms \citep{sutton2000policy, kakade2002natural, schulman2015trust, schulman2017proximal, mnih2016asynchronous, sutton2018reinforcement}, e.g., as seen in stable-baselines3 \citep{raffin2021stable}. 

We trained the PPO algorithm over 100 iterations, using the parameters detailed in the code snippet in Figure \ref{fig:ppo_he1_example}. Observations drawn from the graph on the right of Figure \ref{fig:ppo_he1_example} reveal that the PPO controller, upon convergence, successfully steers three of the four state variables towards zero. However, one state variable settles at approximately $-0.1$, deviating from the target value, which is not ideal. We provide additional examples of applying PPO to a PDE environment and RL algorithms from stable-baselines3 to a linear control environment in our GitHub repository.

\subsection{Generate Uncontrolled PDE Trajectories}\label{sec:zero_control}
In Figure \ref{fig:zero_controller}, we show how to generate uncontrolled PDE trajectories (Figures \ref{fig:cdr}-\ref{fig:ginzburg_landau}) using a zero controller, exemplified through the CDR equation environment. This approach is instrumental for exploring the open-loop dynamics of PDEs, particularly in tuning physical parameters (cf., Section \ref{sec:pde_eigens}) and testing various initial conditions to identify the optimal experimental settings. 

\begin{figure}[t]
\centering
\hspace{0.5em}
\begin{lstlisting}[language=Python, style=mystyle]
import controlgym as gym

if __name__ == "__main__":
	env = gym.make("convection_diffusion_reaction")
	zero = gym.controllers.Zero(env)
	zero.run()
	gym.save(zero)
\end{lstlisting}
\caption{Applying a zero controller to the CDR environment to evaluate the open-loop trajectory}
\label{fig:zero_controller}
\end{figure}

\section{Conclusion}
	We have presented \texttt{controlgym}, a library designed to support the research efforts of L4DC. The \texttt{controlgym} project facilitates a deeper investigation into the performance of RL algorithms, particularly focusing on their convergence, the stability and robustness of RL-based controllers, and the scalability of RL algorithms to systems with high and infinite state dimensionality. 
		
\acks{The research of XZ, WM, and TB were supported in part by the US Army Research Laboratory (ARL) Cooperative Agreement W911NF-17-2-0181, in part by the Army Research Office (ARO) MURI Grant AG285, and in part by the ARO Grant W911NF-24-1-0085. SM and MB were supported solely by MERL.}

\bibliography{main}

\end{document}